\documentclass[pdftex,twocolumn,epjc3]{svjour3}          

\RequirePackage[T1]{fontenc}
\smartqed  

\RequirePackage{graphicx}
\RequirePackage{mathptmx}      
\RequirePackage{flushend}
\RequirePackage[numbers,sort&compress]{natbib}
\RequirePackage[colorlinks,citecolor=blue,urlcolor=blue,linkcolor=blue]{hyperref}

\journalname{Eur. Phys. J. C}

\begin{document}

\title{A comparative study between EGB gravity and GTR by modelling compact stars}
\author{Piyali Bhar\thanksref{e1,addr1} \and Megan Govender\thanksref{e2,addr2} \and Ranjan Sharma\thanksref{e3,addr3}}
\thankstext{e1}{e-mail:piyalibhar90@gmail.com }
\thankstext{e2}{e-mail:megandhreng@dut.ac.za}
\thankstext{e3}{e-mail:rsharma@associates.iucaa.in}

\institute{Department of Mathematics,Government General Degree College, Singur, Hooghly 712 409, West Bengal,
India\label{addr1}
\and
Department of Mathematics, Faculty of Applied Sciences, Durban University of Technology, Durban, 4000, South Africa\label{addr2}
\and
Department of Physics, P. D. Women's College, Jalpaiguri 735101, India.\label{addr3}}

\date{Received: date / Accepted: date}

\maketitle

\begin{abstract}

In this paper we utilise the Krori-Barua ansatz to model compact stars within the framework of Einstein-Gauss-Bonnet (EGB) gravity. The thrust of our investigation is to carry out a comparative analysis of the physical properties of our models in EGB and classical general relativity theory.

\end{abstract}


\section{Introduction}

Modeling stars within the framework of Einstein theory of gravity has occupied researchers for a century. The first exact solution of the Einstein field equations was obtained by Schwarzschild in 1916. This solution describes the exterior vacuum spacetime of a spherically symmetric body. In the same year, Schwarzschild presented the interior solution describing the gravitational behaviour of a uniform density sphere. Since then, hundreds of exact solutions of the classical field equations describing static fluid spheres have been obtained in which the energy-momentum tensor describing the matter distribution incorporated anisotropy, bulk viscosity, electromagnetic field, scalar field, dark energy as well as the cosmological constant. Various techniques ranging from adhoc assumptions of the gravitational potentials, specifying an equation of state, spacetime symmetry, conformal flatness, to name a few, have been employed to solve the field equations. A systematic study of the physical viability of these solutions have been carried out by \citet{Delgaty98} who showed that only a few class of available solutions are capable of describing realistic stellar configurations. Many of these results have later been extended to higher dimensions as well. Interestingly, the dimensionality of spacetime apparently influence the stability of these fluid spheres. With astronomical observations of compact objects becoming more precise and data sets of neutron stars and strange stars being readily available, a new and invigorated search for exact solutions of the $4$-D Einstein field equations are being carried out. In the recent past, there has been an explosion of such solutions describing compact objects which adequately fit observations. Observational data on mass-to-radius ratio, redshift and luminosity profiles are some of the key characteristics for testing the physical validity of these models. One of the objectives of these models is to fine-tune the equation of state in the high density regime. Apart from classical barotropic equation of state $p = \alpha \rho$, many models are being developed based on our current understanding of particle physics. The MIT bag model first proposed by \citet{Chodos74} has been widely used in modeling `strange stars' composed of $u$, $d$ and $s$ quarks. The Chaplygin equation of state \citep{Chap} incorporating dark energy into the stellar fluid configuration has been employed to model compact objects ranging from quark stars through to neutron stars. While these models are successful in accounting for observational data of compact objects, one cannot ignore the fact that gravitational behaviour (metric functions) and thermodynamical behaviour (energy-momentum tensor) can be tweaked by hand to align them with observations. Therefore, the fundamental question to be asked is: whether classical general relativity is sufficient to account for observed stellar characteristics. In other words, can certain, if not all, stellar features reside in higher order theories of gravity?\par

The behaviour and dynamics of the gravitational field can be extended to higher dimensions in a natural way. An elegant and fruitful generalization of classical general relativity is the so-called Einstein-Gauss-Bonnet (EGB) gravity which arises from the incorporation of an additional term to the standard Einstein-Hilbert action, which is quadratic in the Riemann tensor. Varying this additional term with respect to the metric tensor only produces a system of second order equations which are compatible with classical general relativity. In standard $4$-D, EGB and Einstein gravity are indistinguishable. The departure from the standard $4$-D Einstein gravity occurs in higher dimensions. There have been many interesting results in the $5$-D EGB theory ranging from the vacuum exterior solution due to \citet{dg} through to generalization of the Kerr-Schild vacuum solution. The dynamics of gravitational collapse and the resulting end-states in EGB gravity has also received widespread attention. The study of a spherically symmetric inhomogeneous dust (as well as null dust) in EGB gravity with $\alpha > 0$ was shown to alter the causal structure of the singularities compared to the standard $5$-D general relativistic case. The result is, in fact,  a counter example of the cosmic censorship conjecture. The study of Vaidya radiating black-holes in EGB gravity has revealed that the location of the horizons is changed as compared to the standard $4$-D gravity. The universality of Schwarzschild's uniform density solution was established using EGB gravity and later extended to Lovelock gravity. The Buchdahl inequality for static spheres has been extended to $5$-D EGB gravity. It was shown that the sign of the Einstein-Gauss-Bonnet coupling constant plays a crucial role on the mass-to-radius ratio. An interesting outcome of the investigation was that one could pack in more mass in $5$-D EGB compared to standard $4$-D Einstein gravity to achieve stability. Despite the non-linearity of the field equations, several exact solutions in $5$-D EGB gravity have recently been found. The classic isothermal sphere has been generalized to $5$-D EGB gravity. Just as in the $4$-D case, the $5$-D EGB models exhibit a linear barotropic equation of state.\par

In $4$-D gravity, one of the exact solutions which has got much attention is the \citet{kb} solution. It is a solution of the Einstein-Maxwell system describing a spherically symmetric charged fluid sphere. The gravitational potentials are finite everywhere within the stellar distribution and the matter variables are well-behaved \cite{Ivanov2002}. Consequently, the Krori-Barua (KB) solution has been used by many to model compact objects within the framework of Einstein's gravity. Several researchers have utilized various equations of state, ranging from the MIT bag model through to the generalized Chaplygin gas together with the KB ansatz to model compact stars such as  $Her X-1$, $4U 1820-30$, $SAX J 1808.4-3658$, $4U 1728-34$, $PSR 0943+10$ and $RX J185635-3754$ \cite{Junevicus1976,Varela2010,Rahaman2010,Rahaman2012}. In this work, we intend to extend the KB solution to $5$-D EGB gravity. The motivation for this modification is to analyze the effects, if any, of the EGB term on stability, compactness and other physical features of compact stellar objects.

\section{Einstein-Gauss-Bonnet Gravity}\label{sec:2}
The Gauss-Bonnet action in five dimensions is written as
\begin{equation}\label{1}
S=\int \sqrt{-g} \left[\frac{1}{2}(R-2 \Lambda +\alpha
L_{GB})\right]d^5x +S_{matter},
\end{equation}
where $\alpha$ is the Gauss-Bonnet coupling constant. The strength of the action $L_{GB}$ lies in the fact that despite the
Lagrangian being quadratic in the Ricci tensor, Ricci scalar and the Riemann tensor, the equations of motion turn out to be second
order quasi-linear which are compatible with the standard Einstein formalism of gravity. The Gauss-Bonnet term is of no consequence for $n\leq 4$ but is
generally nonzero for $n \geq 5$.

The EGB field equations may be written as
\begin{equation}\label{2}
G_{ab}+\alpha H_{ab}=T_{ab},
\end{equation}
with metric signature $(- + + + +)$ where $G_{ab}$ is the Einstein
tensor. The coupling constant $\alpha$ is related to the inverse string tension arising from the low energy effective action of string theory and to this end we consider $\alpha \geq 0$. The Lanczos tensor is given by
\begin{eqnarray}\label{3}
H_{ab}&= &2\left(R R_{ab}-2R_{ac}R^c_b- 2
R^{cd}R_{acbd}+R^{cde}_aR_{bcde}\right)\nonumber\\
&&-\frac{1}{2}g_{ab}L_{GB},
\end{eqnarray}
where the Lovelock term has the form

\begin{equation}\label{4}
L_{GB}=R^2 +R_{abcd}R^{abcd}- 4R_{cd}R^{cd}.
\end{equation}
In the above formalism we use geometric units with the coupling constant $\kappa$ set to unity.

\section{Field equations}
\label{sec:3}The $5$-dimensional line element for a static spherically symmetric spacetime has the standard form
\begin{eqnarray}
\label{5} ds^{2}& =& -e^{2\nu(r)} dt^{2} + e^{2\lambda(r)} dr^{2} +
 r^{2}(d\theta^{2} + \sin^{2}{\theta} d\phi^2\nonumber\\
 && +\sin^{2}{\theta} \sin^{2}{\phi^2}
 d\psi),
\end{eqnarray}
in coordinates ($x^i = t,r,\theta,\phi,\psi$). By considering the comoving fluid velocity as $u^a=e^{-\nu}\delta_0^a$, the EGB field equation (\ref{2}) yields the following set of independent equations
\begin{eqnarray}
\label{7a}\rho&=& \frac{3}{e^{4\lambda }r^3} \left(4\alpha
\lambda'
+re^{2\lambda}-re^{4\lambda}- r^2 e^{2\lambda}\lambda' -4\alpha e^{2\lambda}\lambda'\right),\\
\label{7b} p_r& = &\frac{3}{e^{4\lambda }r^3}
\left[-re^{4\lambda}+
(r^2 \nu' +r +4\alpha \nu')e^{2\lambda} -4\alpha \nu'\right] , \\
\label{7c} p_t&=& \frac{1}{e^{4\lambda }r^2} \left(- e^{4\lambda
}- 4\alpha \nu''+ 12 \alpha \nu' \lambda' -4 \alpha
(\nu')^2\right)
\nonumber\\
&& +\frac{1}{e^{2\lambda }r^2} \left(1- r^2 \nu' \lambda' +2r \nu'
-2r  \lambda' +r^2(\nu')^2 \right) \nonumber \\
&& +\frac{1}{e^{2\lambda }r^2} \left(r^2 \nu'' -4\alpha
\nu'\lambda' + 4\alpha (\nu')^2 +4\alpha \nu''\right),
\end{eqnarray}
where $\rho$, $p_r$ and $p_t$ respectively denote the matter density, radial and transverse pressure of the fluid. Note that a $\prime$ denotes the differentiation with respect to the radial coordinate $r$.

\section{A particular solution}
Note that equations (6)-(8) correspond to a system of three linearly independent equations with five unknowns, namely $\rho$, $p_r$, $p_t$, $\lambda$ and $\nu$. To analyze behaviour of the physical parameters, we assume that the metric potentials are given by the \citet{kb} solution
\[2\lambda(r) = Ar^{2}, 2\nu(r) = Br^{2}+C\]
where $A$, $B$ and $C$ are undetermined constants which can be obtained from the matching conditions.

Consequently, the matter density and the two pressures are obtained as
\begin{eqnarray}
\rho &=& \frac{3e^{-2Ar^{2}}}{r^{2}}\left[-4A\alpha+e^{2Ar^{2}}+e^{Ar^{2}}\left(4A\alpha+Ar^{2}-1\right)\right]\label{den},\\
 p_r &=& \frac{3e^{-2Ar^{2}}}{r^{2}}\left[-4\alpha B-e^{2Ar^{2}}+e^{Ar^{2}}\left(1+4\alpha B+Br^{2}\right)\right]\label{pr},\\
p_t &=&\frac{e^{-2Ar^{2}}}{r^{2}}\left[-e^{-2Ar^{2}}-4\alpha B\left\{1+(B-3A)r^{2}\right\}\right]\nonumber\\
&&+\frac{e^{-Ar^{2}}}{r^{2}}\left[1+4\alpha B+\left\{B(3+4\alpha B)-2A(1+2\alpha B) \right\}r^{2}\right.\nonumber\\
&&\left.+B(B-A)r^{4}\right]\label{pt}.
\end{eqnarray}
The anisotropy $\Delta=p_t-p_r$ is obtained as

\begin{eqnarray}
\Delta&=&\frac{e^{-2Ar^{2}}}{r^{2}}\left[2e^{2Ar^{2}}+4\alpha B \left\{2+(3A-B)r^{2}\right\}\right]\nonumber\\
&&-\frac{e^{-Ar^{2}}}{r^{2}}\left[2+2Ar^{2}+(A-B)Br^{4}\right.\nonumber\\
&&\left.+4\alpha B \left\{2+(A-B)r^{2}\right\}\right].
\end{eqnarray}

\section{Exterior spacetime and matching conditions}
The static exterior spacetime in $5$-D is described by the Einstein-Gauss-Bonnet-Schwarzschild
solution\cite{dg}
\begin{eqnarray}
\label{8} ds^{2}& =& -F(r) dt^{2} + [F(r)]^{-1} dr^{2} +
 r^{2}\left(d\theta^{2} + \sin^{2}{\theta} d\phi^2\right.\nonumber\\
 &&\left. +\sin^{2}{\theta} \sin^{2}{\phi}
 d\psi\right),
\end{eqnarray}
where,
\begin{equation}
\label{9} F(r) =1+\frac{r^2}{4 \alpha}
\left(1-\sqrt{1+\frac{8\alpha M}{r^4}}\right).
\end{equation}
In (\ref{9}) $M$ is associated with the gravitational mass of the hypersphere.

Using continuity of the metric functions and their derivatives, namely $g_{rr}$, $g_{tt}$ and $\frac{\partial g_{tt}}{\partial r}$ across the boundary $r=R$ we get,
\[e^{-AR^{2}}=1+\frac{R^2}{4 \alpha}
\left(1-\sqrt{1+\frac{8\alpha M}{R^4}}\right),\]
\[e^{BR^{2}+C}=1+\frac{R^2}{4 \alpha}
\left(1-\sqrt{1+\frac{8\alpha M}{R^4}}\right),\]
\[2B e^{BR^{2}+C}=-\frac{1}{2\alpha}\frac{1-\sqrt{1+\frac{8\alpha M}{R^{4}}}}{\sqrt{1+\frac{8\alpha M}{R^{4}}}}.\]
Solving the above set of equations we get,
\begin{equation}
A=-\frac{1}{R^{2}}\ln\left[1+\frac{R^{2}}{4\alpha}\left(1-\sqrt{1+\frac{8\alpha R}{R^{4}}}\right)\right],
\end{equation}
\begin{equation}
B=-\frac{1}{4\alpha}\frac{1-\sqrt{1+\frac{8\alpha M}{R^{4}}}}{\sqrt{1+\frac{8\alpha M}{R^{4}}}}\frac{1}{1+\frac{R^{2}}{4\alpha}\left(1-\sqrt{1+\frac{8\alpha R}{R^{4}}}\right)},
\end{equation}
\begin{equation}
C=\ln\left[1+\frac{R^{2}}{4\alpha}\left(1-\sqrt{1+\frac{8\alpha R}{R^{4}}}\right)\right]-BR^{2}.
\end{equation}

\section{Physical features}

Physical features of our model are outlined below:
\begin{enumerate}
\item For a physically acceptable model, the energy density $\rho$ and two pressures $p_r$ and $p_t$  should be positive inside the star. Also, the radial pressure must vanish at a finite radial distance while the tangential pressure $p_t$  need not vanish at the boundary.

In our model, we note that for specific choices of the model parameters (we have assumed $A=0.006$, $B=0.008$ and $\alpha=0,~1.5$), the density and two pressures remain positive throughout the stellar interior (see Fig.~\ref{fg1}, Fig.~\ref{fg2} and Fig.~\ref{fg3} ). Most interestingly, the radial pressure $p_r$ vanishes at a greater radial distance ($10~$km) in EGB gravity as compared to GTR ($9.58~$km).
\begin{figure}[htbp]
    \centering
        \includegraphics[scale=.3]{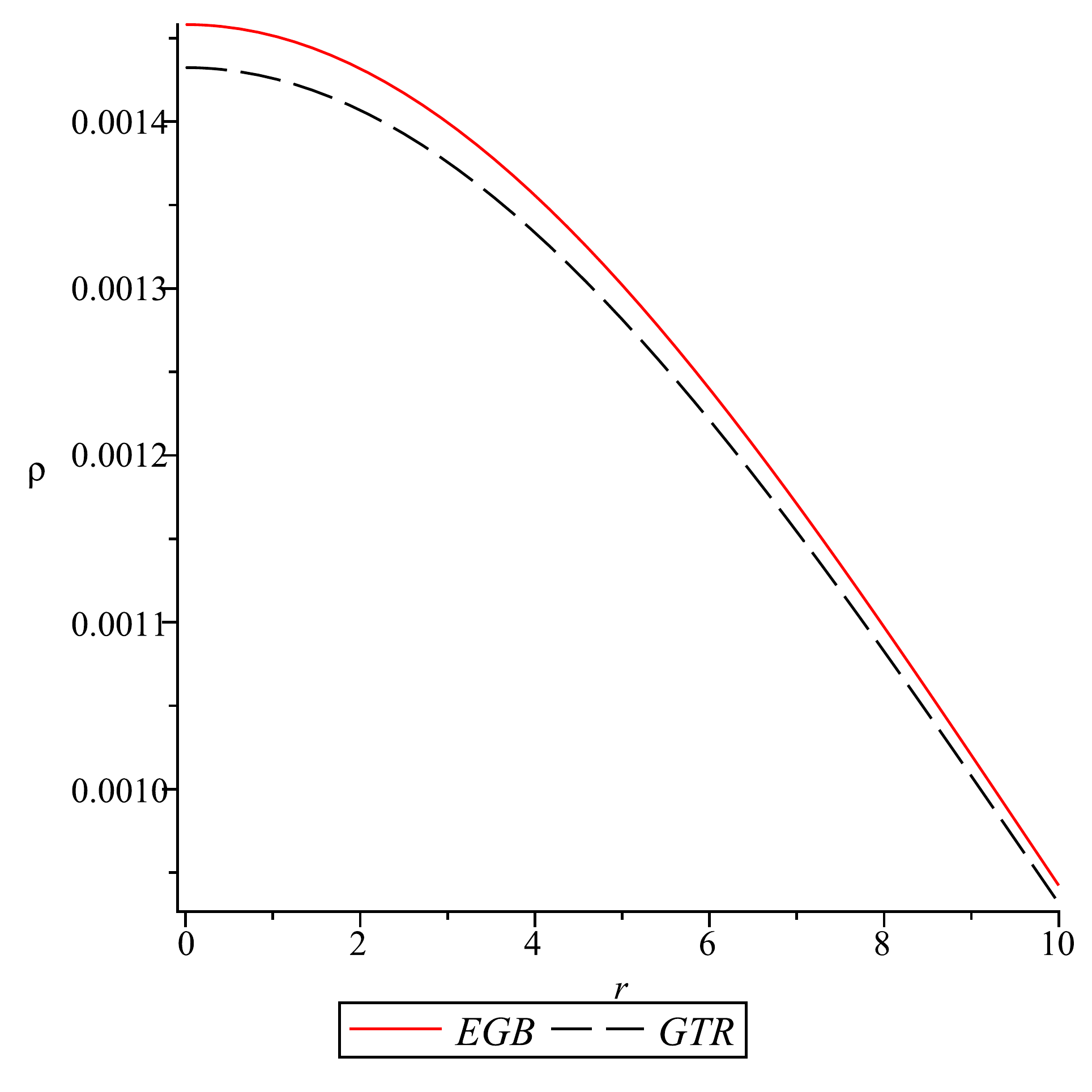}
       \caption{Matter density $\rho$ plotted against the radial distance $r$ where the (red) solid line and the dashed line correspond to EGB gravity and GTR, respectively. We have considered two cases: (i) $A=0.006$, $B=0.008$ and $\alpha=1.5$ (EGB gravity) and (ii) $A=0.006$, $B=0.008$ and $\alpha=0$ (GTR).}
\label{fg1}
\end{figure}

\begin{figure}[htbp]
    \centering
        \includegraphics[scale=.3]{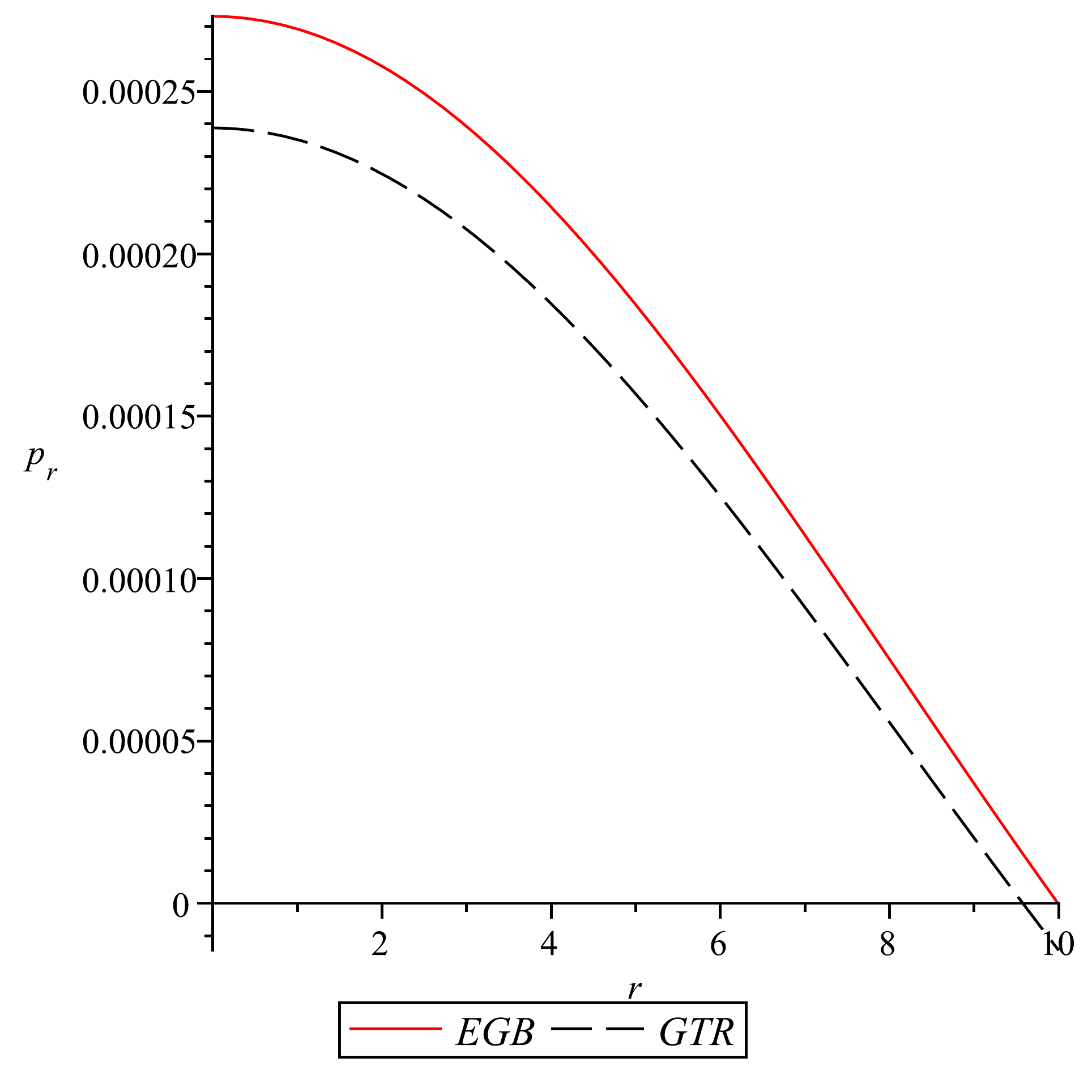}
       \caption{Radial pressure $p_r$ plotted against against the radial distance $r$ by taking the same values of the constants mentioned in fig.\ref{fg1} }
\label{fg2}
\end{figure}


\item Fig.~\ref{fg4} shows that anisotropy is zero at the centre, i.e., $\Delta (r=0) = 0$ and it increase towards the surface. Interestingly, the higher dimensional correctional term has very little or no effect on the anisotropic stress as can be seen in the plot.

\begin{figure}[htbp]
    \centering
        \includegraphics[scale=.3]{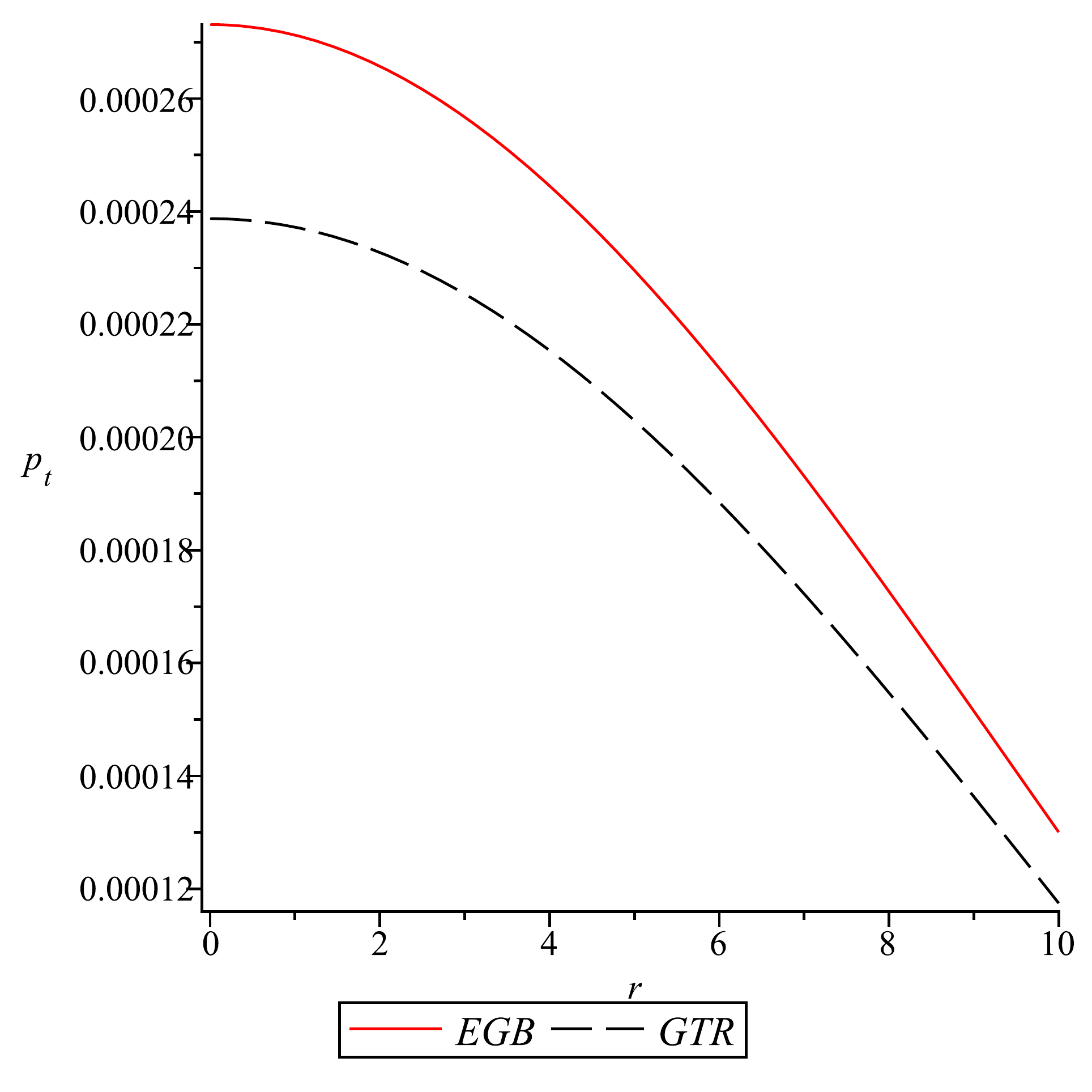}
       \caption{Transverse pressure $p_t$ plotted against against the radial distance $r$ by taking the same values of the constants mentioned in fig.\ref{fg1}}
\label{fg3}
\end{figure}

\begin{figure}[htbp]
    \centering
        \includegraphics[scale=.3]{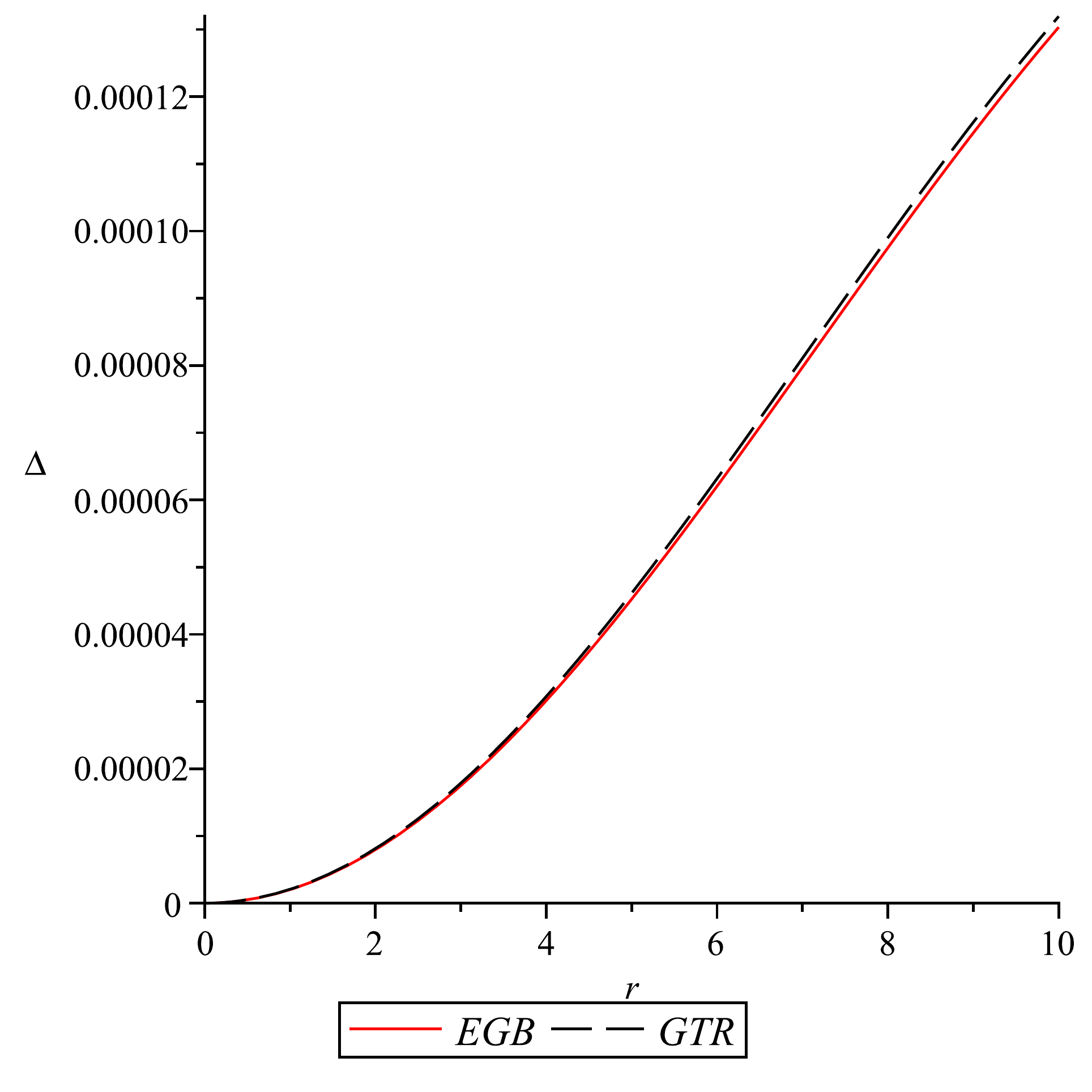}
       \caption{Anisotropic factor $\Delta$ plotted against against the radial distance $r$ by taking the same values of the constants mentioned in fig.\ref{fg1} }
\label{fg4}
\end{figure}


\item From Eqs.~(\ref{den}) and (\ref{pr}), we have
\begin{eqnarray}
\frac{d\rho}{dr}&=&-\frac{3e^{-2Ar^{2}}}{4\pi r^{3}}\left[e^{Ar^{2}}\left(4A^{2}r^{2}\alpha +A^{2}r^{4}-Ar^{2}+4A\alpha-1\right)\right.\nonumber\\
&&\left.-4A\alpha(1+2Ar^{2})+e^{2Ar^{2}}\right],\\
\frac{dp_r}{dr}&=&-\frac{3e^{-2Ar^{2}}}{4\pi r^{3}}\left[e^{Ar^{2}}\left(Ar^{2}+4Ar^{2}\alpha B+ABr^{4}+1+4\alpha B\right)\right.\nonumber\\
&&\left.-4\alpha B(1+2Ar^{2})-e^{2Ar^{2}}\right].
\end{eqnarray}
The radial variation of density and pressure have been shown in Fig.~\ref{fg5}and Fig.~\ref{fg6}respectively. which clearly shows that $\frac{dp_r}{dr} < 0$ and $\frac{d\rho}{dr} < 0$, both in EGB gravity and in GTR.


\begin{figure}[htbp]
    \centering
        \includegraphics[scale=.3]{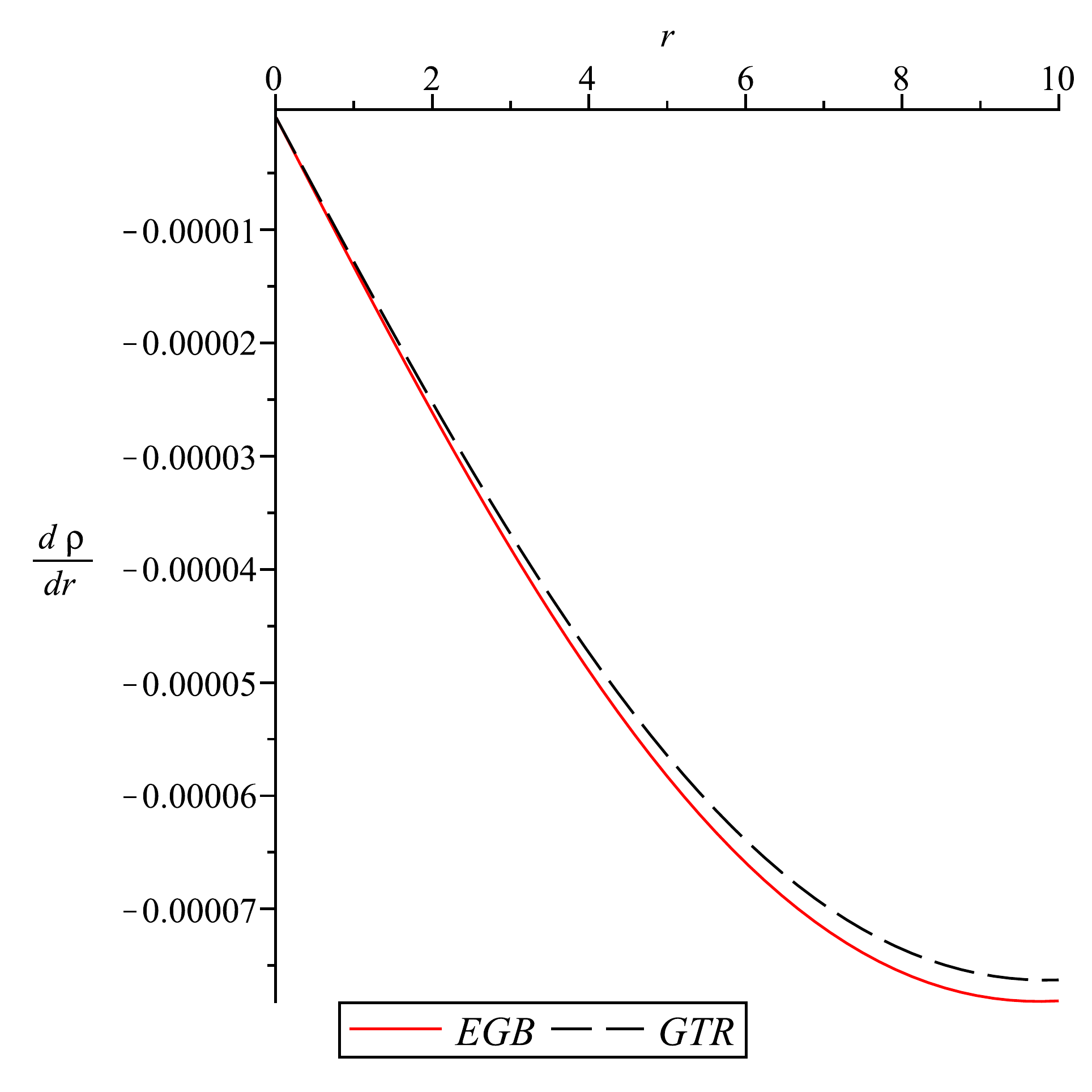}
       \caption{The derivative of the matter density $\rho$ plotted against against the radial distance $r$ by taking the same values of the constants mentioned in fig.\ref{fg1} }
\label{fg5}
\end{figure}

\begin{figure}[htbp]
    \centering
        \includegraphics[scale=.3]{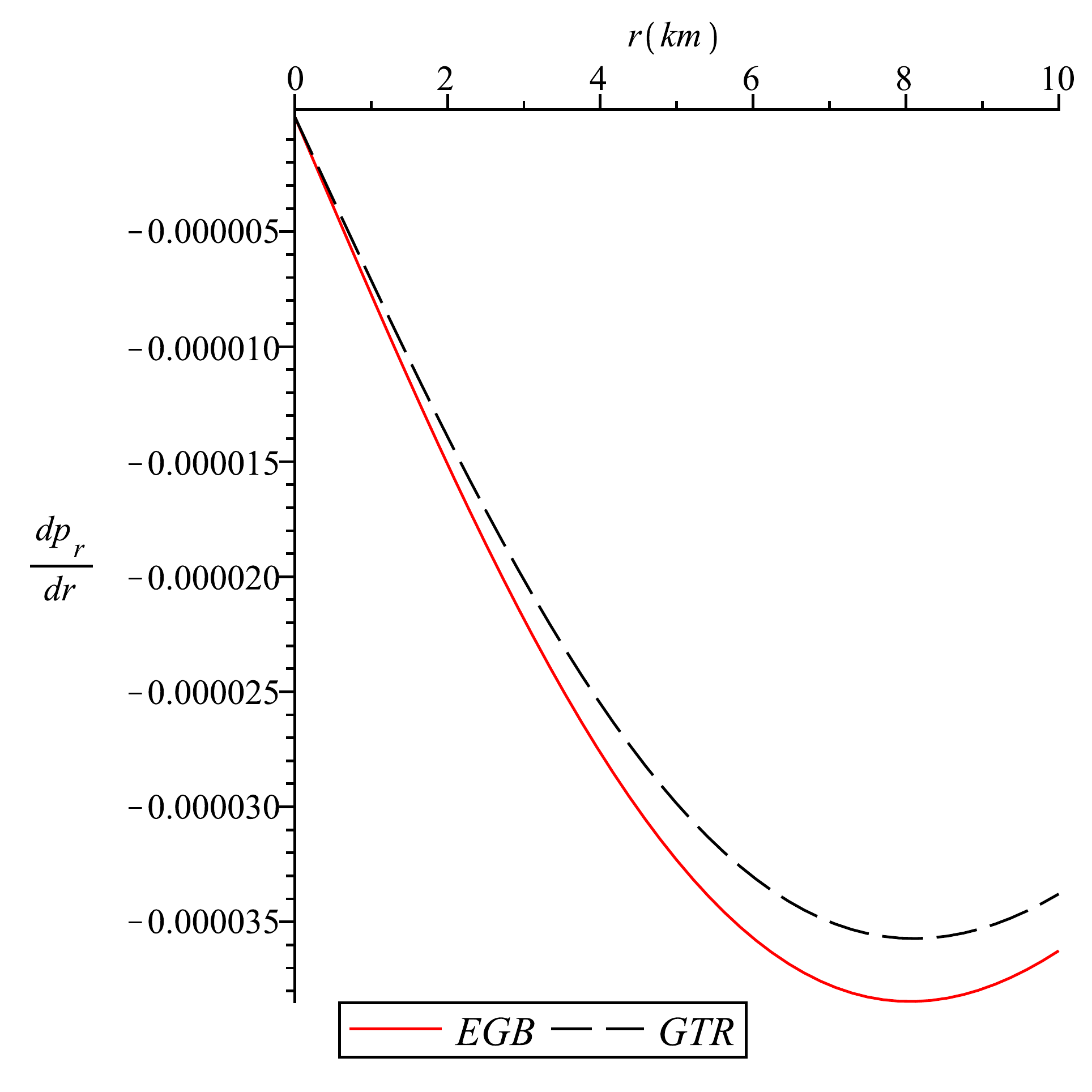}
       \caption{The derivative of radial pressure $p_r$ plotted against against the radial distance $r$ by taking the same values of the constants mentioned in fig.\ref{fg1} }
\label{fg6}
\end{figure}

\item Fig.~\ref{fg7} shows that the ratio of trace of stress tensor to energy density $(p_r+2p_t)/\rho$ decreases
radially outward both in EGB gravity as in GTR which is a desirable feature for a fluid sphere.

\begin{figure}[htbp]
    \centering
        \includegraphics[scale=.3]{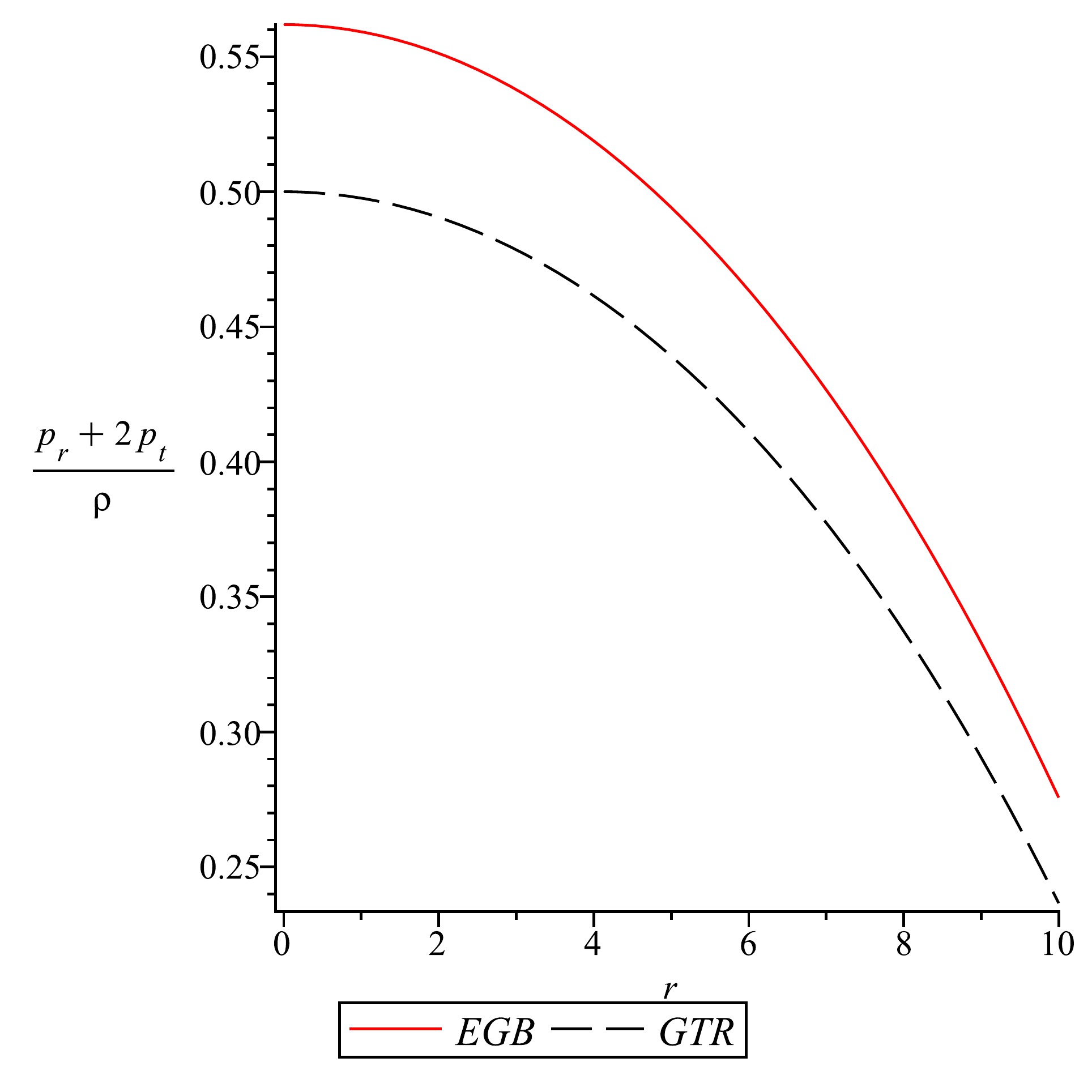}
       \caption{$\frac{p_r+2p_t}{\rho}$ plotted against radial distance $r$ Radial pressure by taking the same values of the constants mentioned in fig.\ref{fg1} }
\label{fg7}
\end{figure}

\item For a physically acceptable model, it is expected that the radial sound speed $v_{sr}^{2} (=\frac{dp_r}{d\rho})$ and transverse sound speed $v_{st}^{2} (=\frac{dp_t}{d\rho})$ should be causal, i.e.,  we should have $ 0 < v_{sr}^{2} \leq 1$ and $ 0 <  v_{st} \leq 1$. We have shown graphically that both in EGB gravity and GTR, the causality condition is not violated at any point within the stellar interior (see Fig.~\ref{fg8} and Fig.~\ref{fg9}). It is evident that in EGB gravity, both the sound speeds take a higher value as compared to its GTR counterpart.

\begin{figure}[htbp]
    \centering
        \includegraphics[scale=.3]{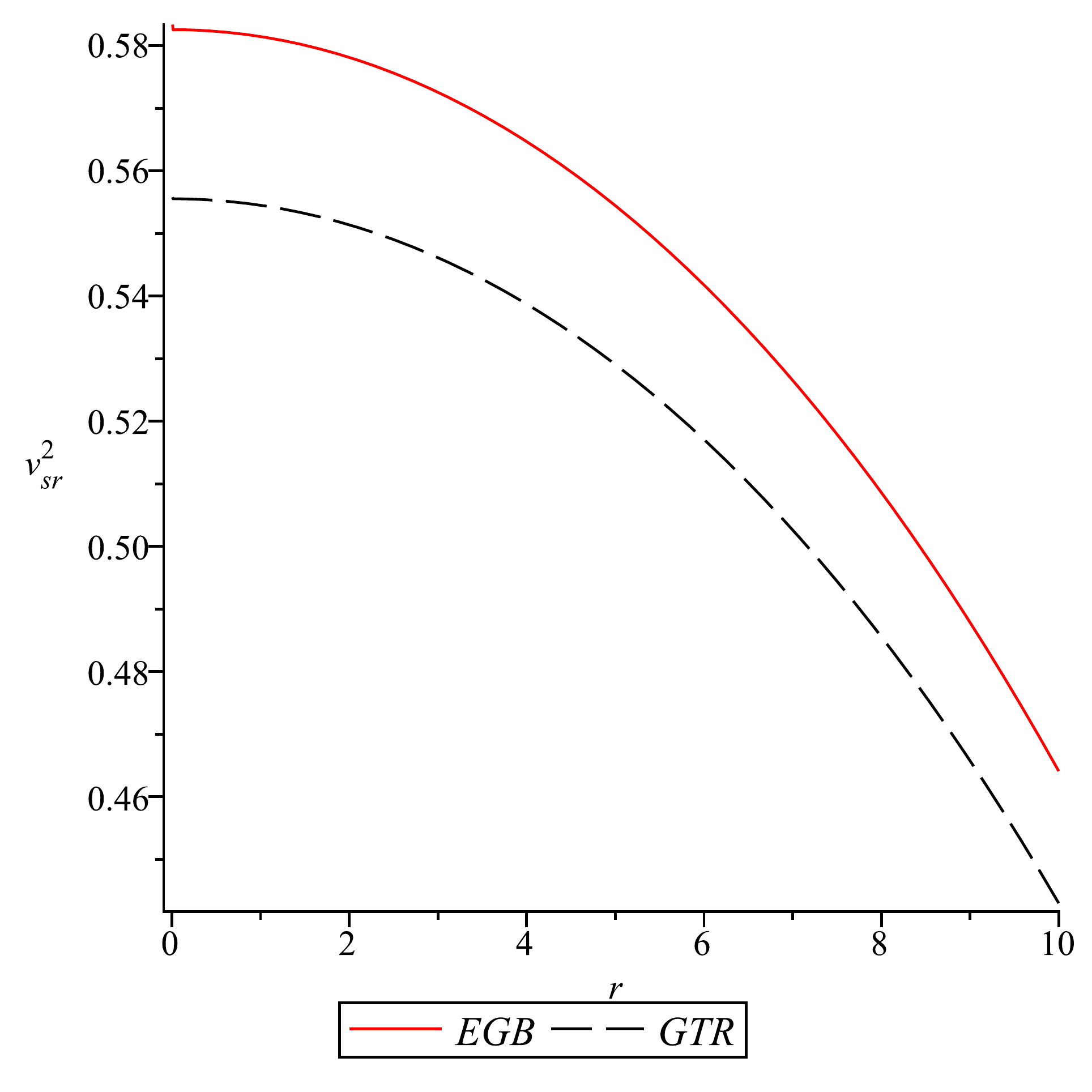}
       \caption{Radial sound velocity $v_{sr}^2$ plotted against against the radial distance $r$ by taking the same values of the constants mentioned in fig.\ref{fg1} }
\label{fg8}
\end{figure}

\begin{figure}[htbp]
    \centering
        \includegraphics[scale=.3]{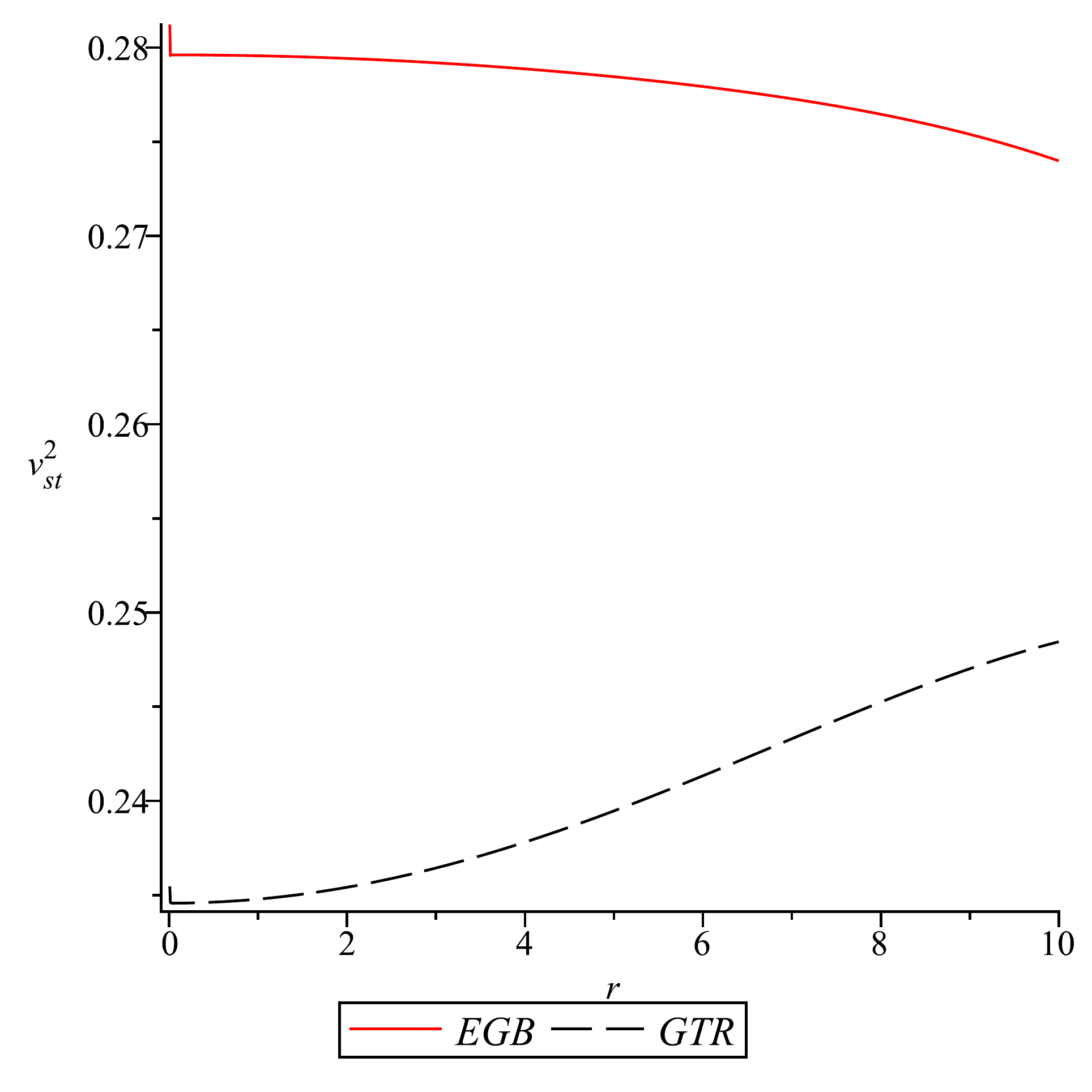}
       \caption{Transverse sound velocity $v_{st}^2$ plotted against against the radial distance $r$ by taking the same values of the constants mentioned in fig.\ref{fg1} }
\label{fg9}
\end{figure}

\end{enumerate}

\subsection{Stability}
\begin{figure}[htbp]
    \centering
        \includegraphics[scale=.3]{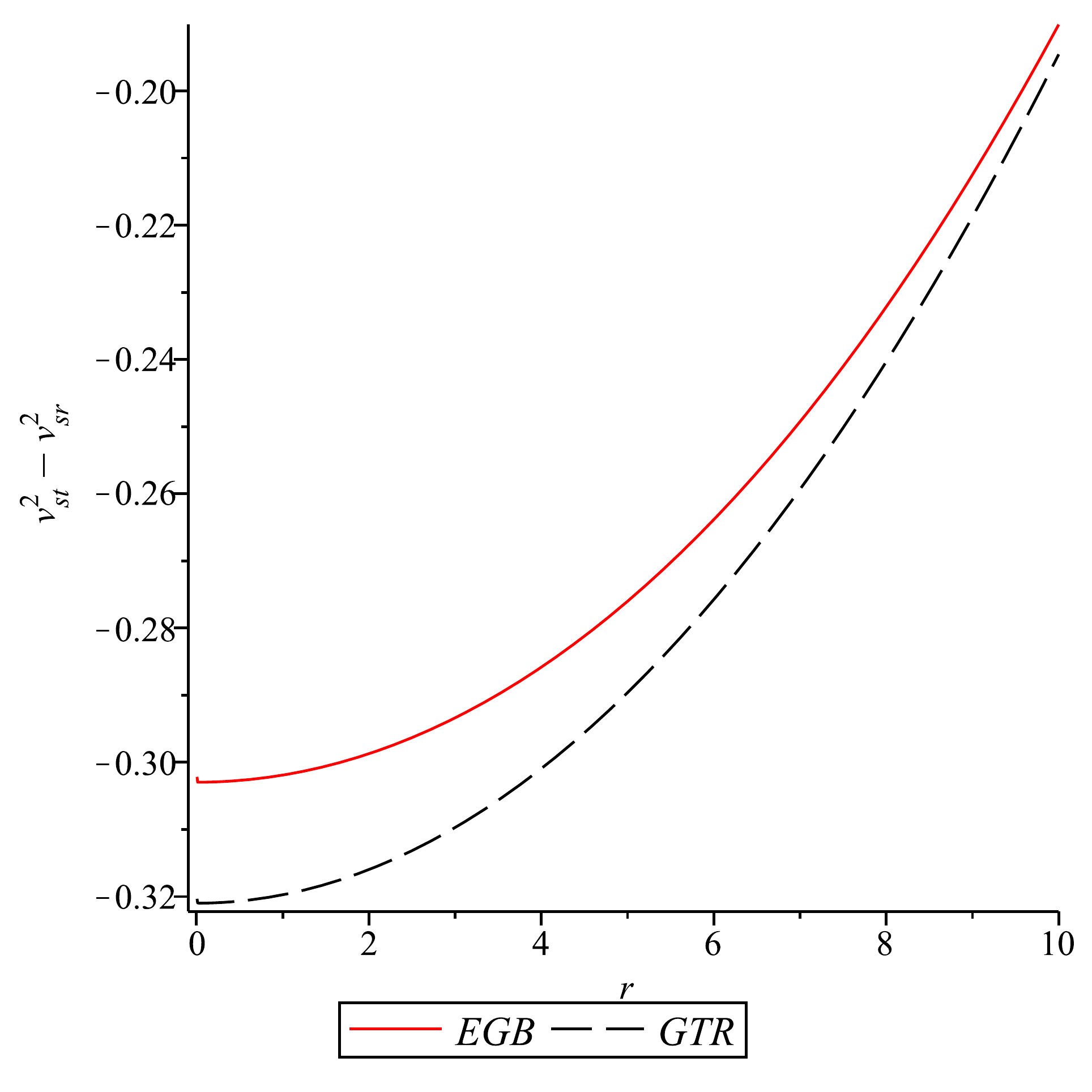}
       \caption{$v_{st}^2-v_{sr}^2$ plotted against against the radial distance $r$ by taking the same values of the constants mentioned in fig.\ref{fg1} }
\label{fg10}
\end{figure}

To check stability of the configuration, we follow the cracking (or overturning) method of Herrera\cite{her} which suggests that a potentially stable region is one where the inequality $ v_{st}^{2} - v_{sr}^{2} < 0$ holds. In Fig.~\ref{fg10}, we have shown the difference of $ v_{st}^{2} -v_{sr}^{2}$ do not change sign which clearly indicates that the configuration is stable for the assumed set of values.

\begin{figure}[htbp]
    \centering
        \includegraphics[scale=.3]{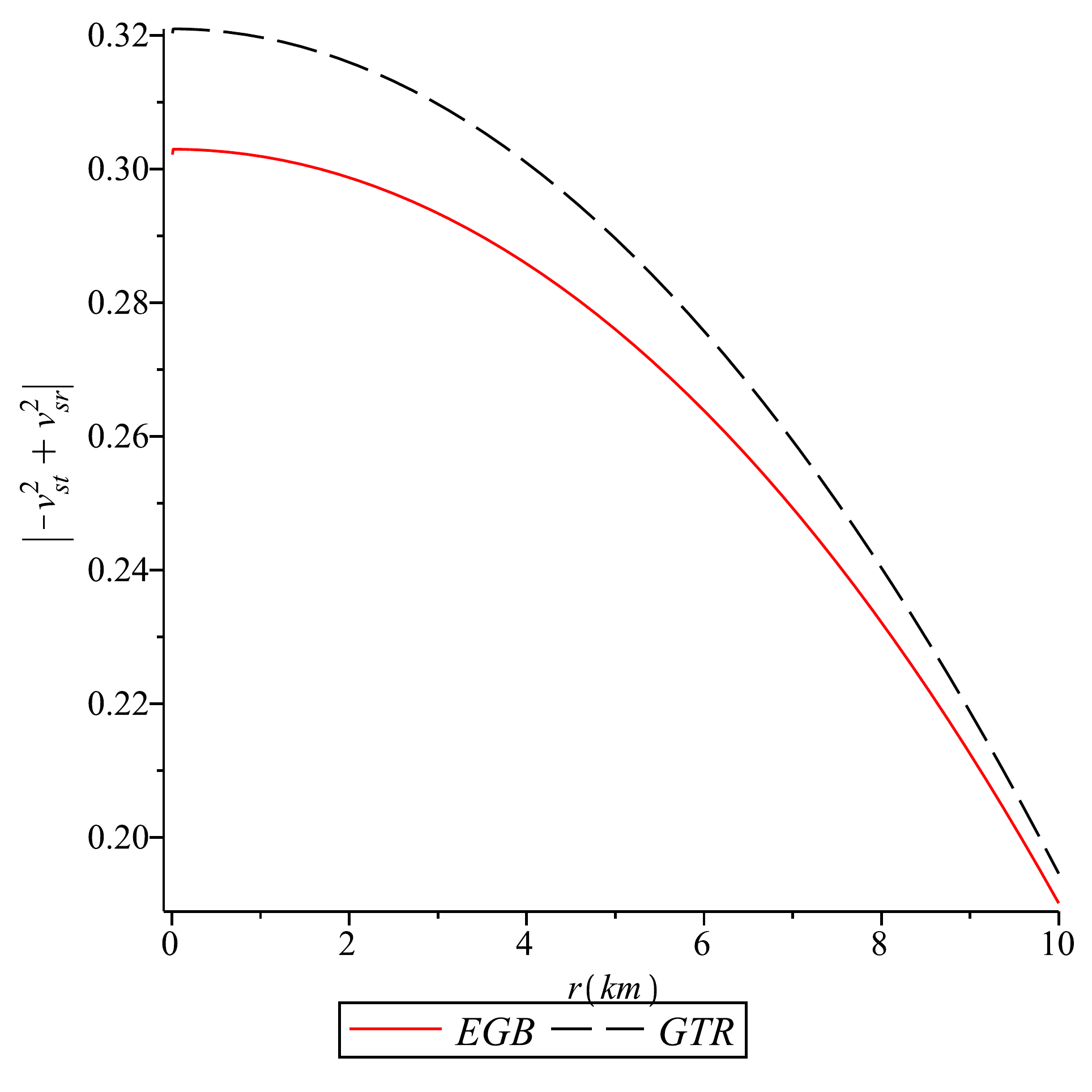}
       \caption{$|v_{st}^2-v_{sr}^2|$ plotted against against the radial distance $r$ by taking the same values of the constants mentioned in fig.\ref{fg1}}
\label{fg11}
\end{figure}

\subsection{Energy Conditions}
\begin{figure}[htbp]
    \centering
        \includegraphics[scale=.3]{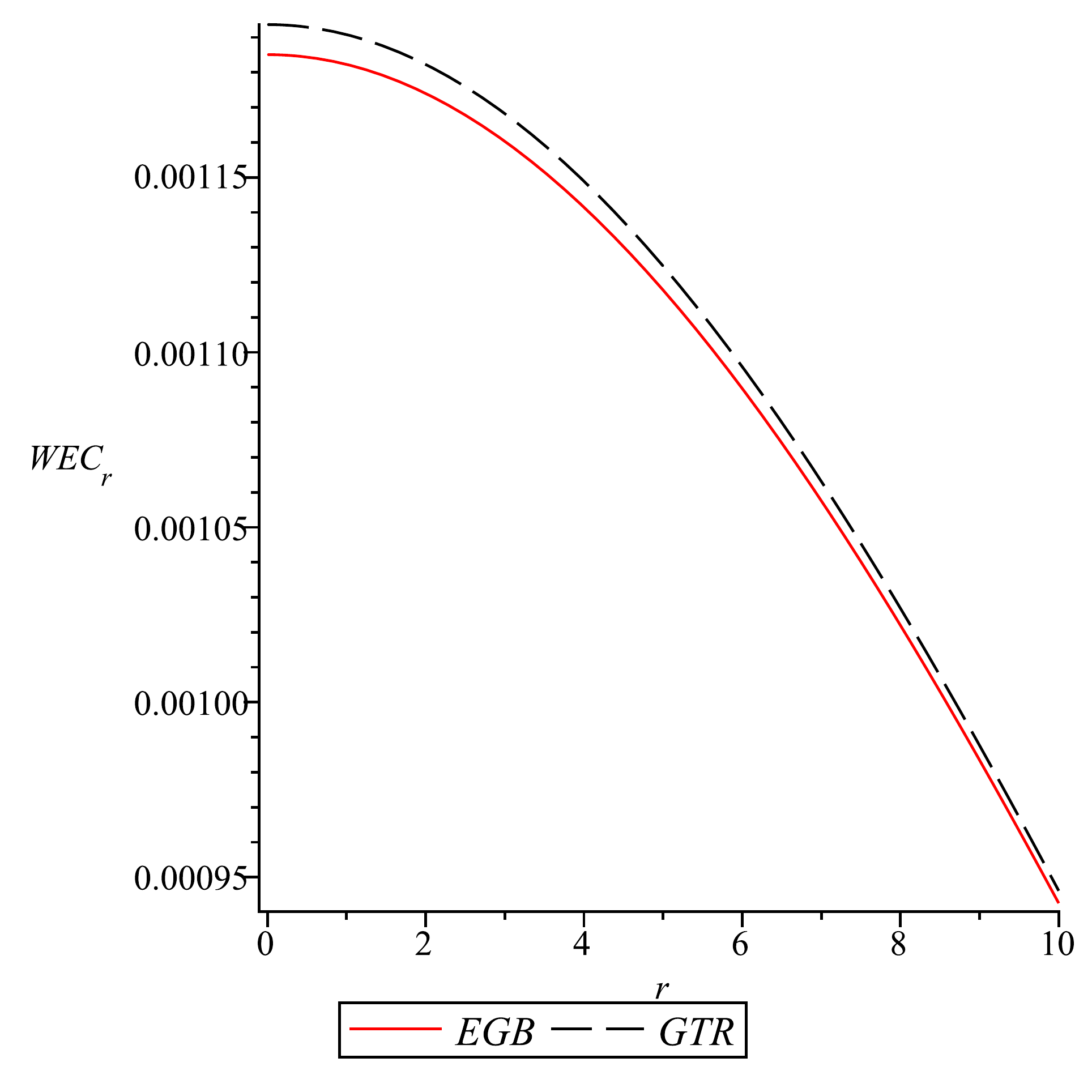}
       \caption{$WEC_r$ plotted against radial distance $r$.}
\label{fg12}
\end{figure}

\begin{figure}[htbp]
    \centering
        \includegraphics[scale=.3]{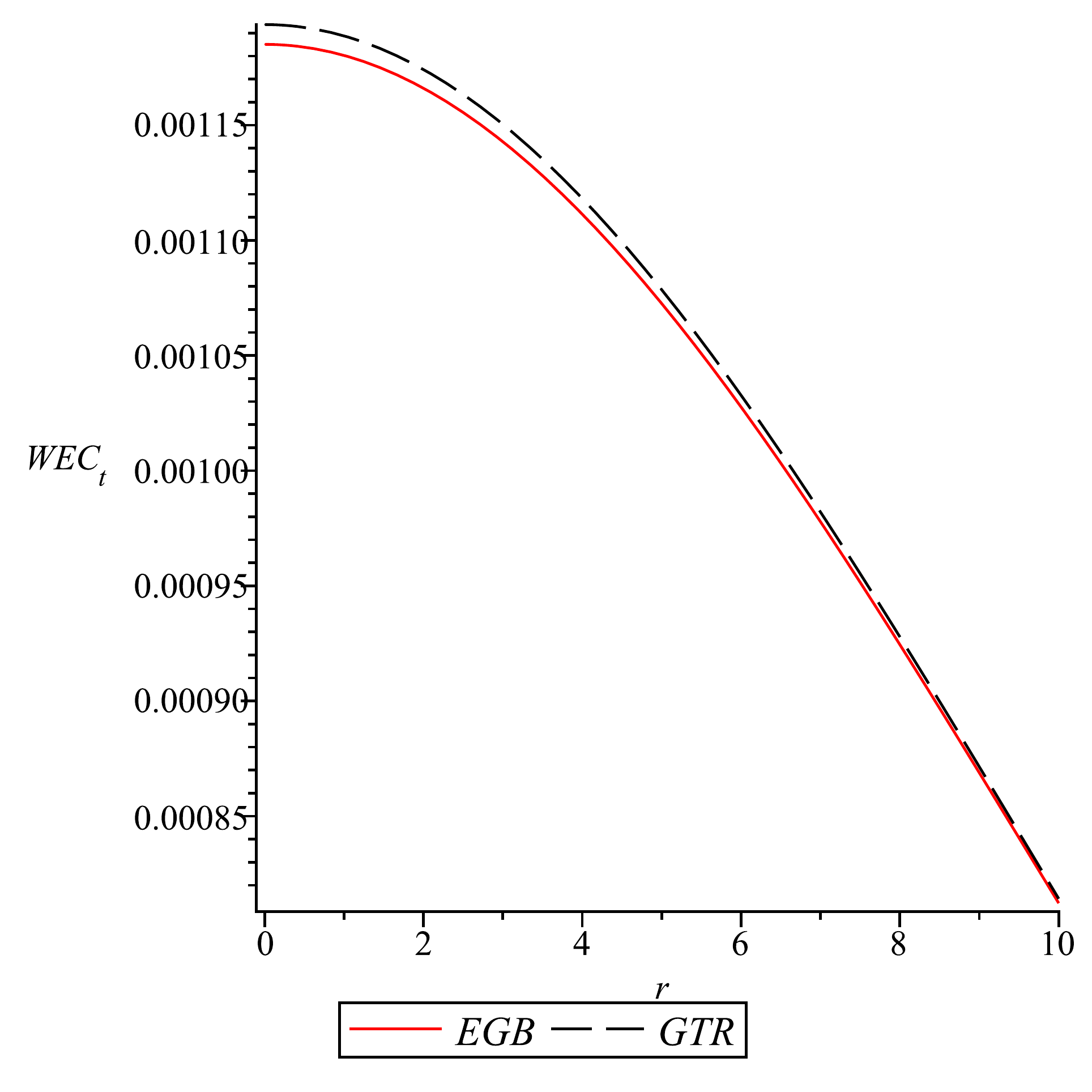}
       \caption{$WEC_t$ plotted against radial distance $r$.}
\label{fg13}
\end{figure}

\begin{figure}[htbp]
    \centering
        \includegraphics[scale=.3]{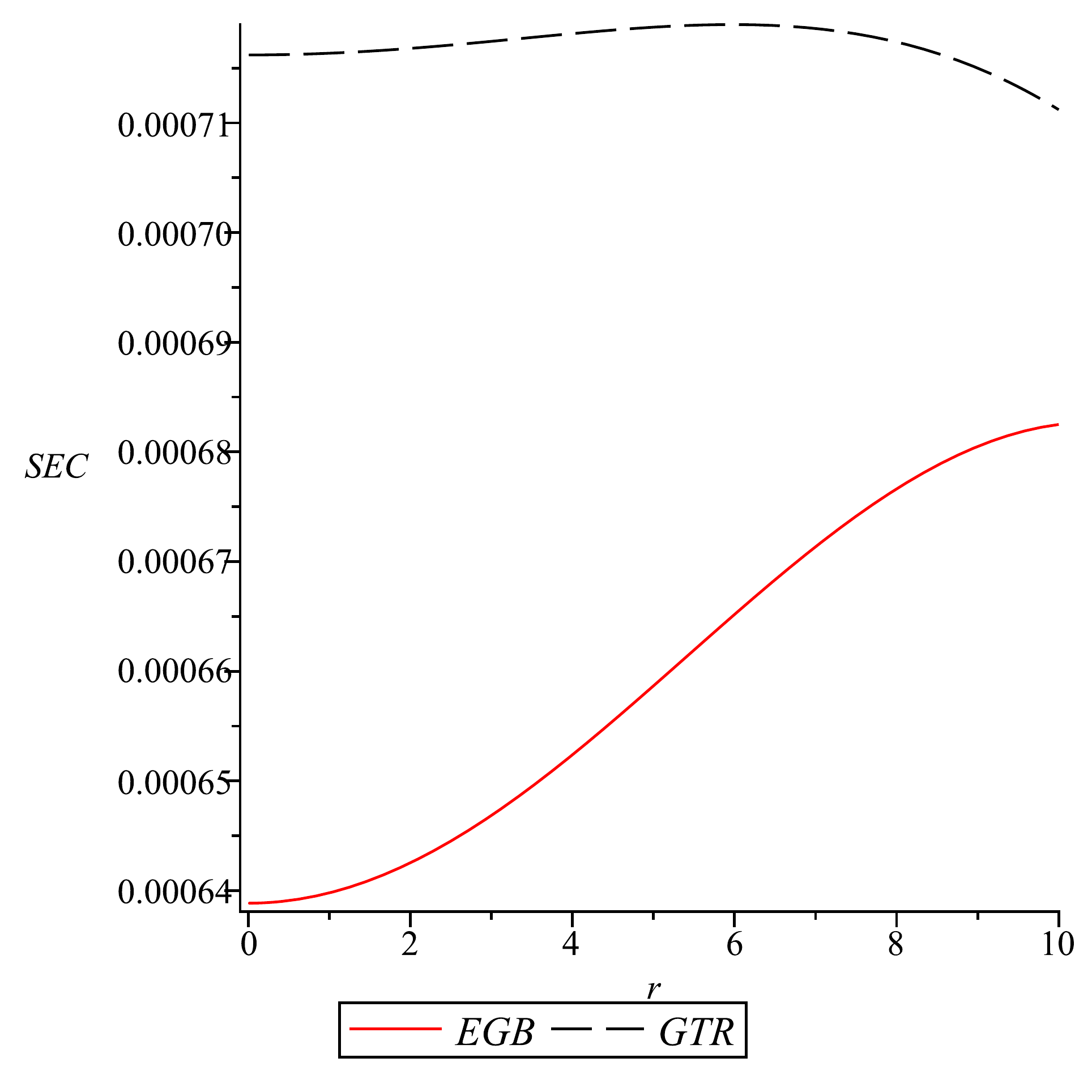}
       \caption{Verification of strong energy condition. }
    \label{fg14}
\end{figure}

Let us now check whether our anisotropic stellar model satisfies the following energy conditions:
\begin{eqnarray}
(i)~ NEC:~~~~\rho\geq 0.\\
(ii)~WEC:~~~~\rho-p_r\geq 0,~~~\rho-p_t \geq 0.\\
(iii)~SEC:~~~~ \rho-p_r-2p_t\geq 0.
\end{eqnarray}
In Fig.~(\ref{fg12}-\ref{fg14}), we have shown graphically that all the energy conditions are satisfied for the assumed set of values of the model parameters.

\subsection{Adiabatic index}
\begin{figure}
    \centering
        \includegraphics[scale=.3]{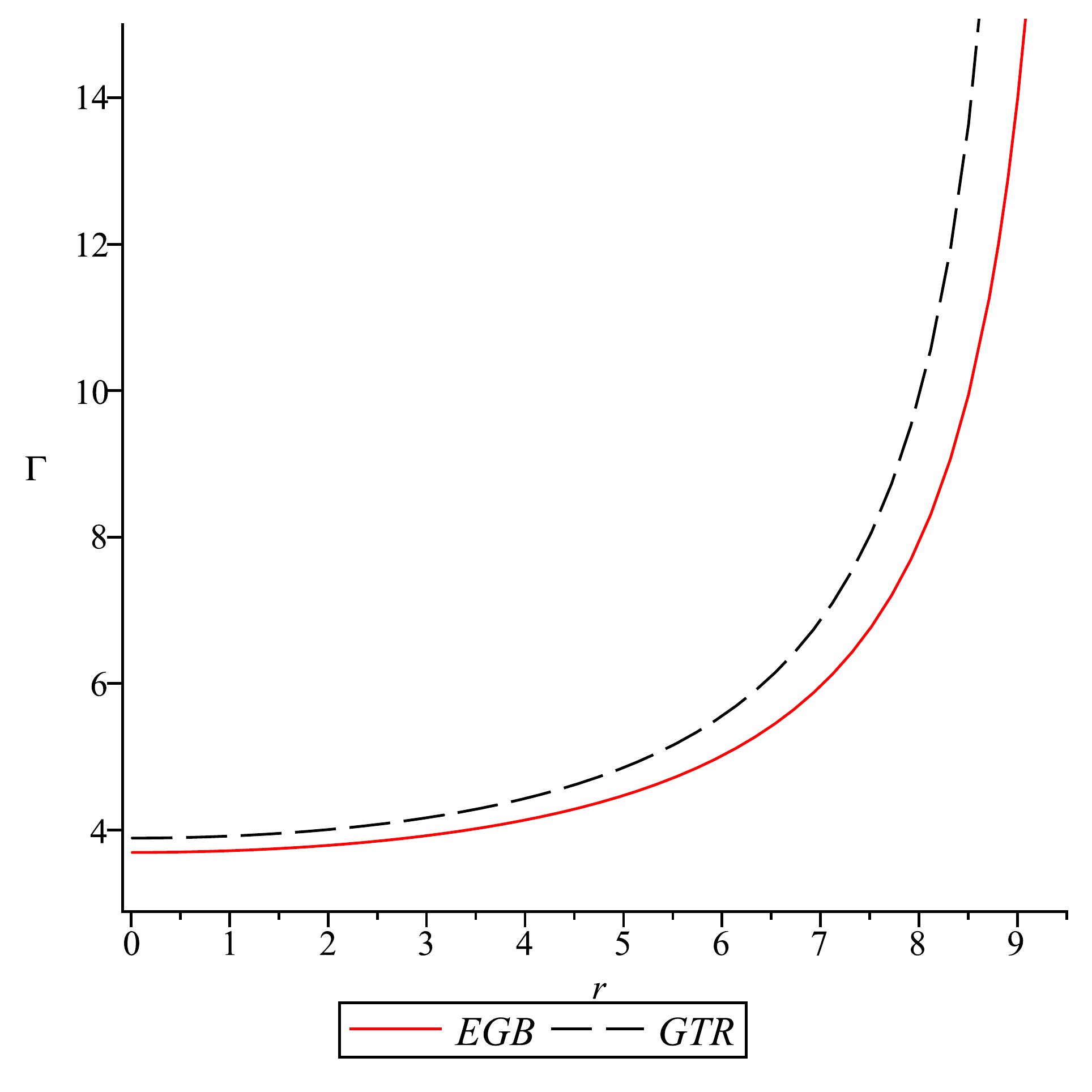}
       \caption{The adiabatic index $\Gamma$ plotted against $r$.}
   \label{fg15}
\end{figure}

It has been proposed by Heintzmann and Hillebrandt \cite{hm} that a neutron star with an anisotropic equation of state would be stable if the adiabatic index $\Gamma > 4/3$. We have evaluated the values of the adiabatic index
\begin{equation}
\Gamma = \frac{\rho+p_r}{p_r}\frac{dp_r}{d\rho},
\end{equation}
both in GTR as well as in EGB gravity. In Fig.~\ref{fg15}, it has been shown that $\Gamma > 4/3$ everywhere within the stellar interior in GRT and in EGB gravity.

\section{Discussion}
Gravitational theories with higher derivative curvature terms developed in the context of string theory in particular, have long been an area of great research attraction. Studies of gravitational behaviour in dimensions $n > 4$ have often been found to yield many non-trivial and interesting results. Of particular interest is the Einstein-Gauss-Bonnet gravity in which the Lagrangian includes a second-order Lovelock
term as the higher curvature correction to GTR. Several vacuum solutions in $5$-dimensional Einstein-Gauss-Bonnet gravity have been found and applied in astrophysics and cosmology. However, it is extremely difficult to generate interior solutions corresponding to star like systems in higher dimensions due to complex nature of the field equations and lack of sufficient information about the equation of state (EOS) of the matter content of the system. In this paper, rather than providing new solutions, we have developed the $4$-dimensional Krori and Barua stellar solution in the context of EGB gravity and analyzed the impacts of the higher derivative correction term on the gross physical behaviour of a relativistic star. Based on physical requirements, bounds on the model parameters have been identified. Within the admissible bounds, for a particular set of values of model parameters, physical characteristics of the developed stellar configuration in EGB gravity have been analyzed in details. In Fig.~\ref{fg1} - \ref{fg15}, graphical representation of various physically meaningful parameters has permitted us to investigate the higher derivative coupling term $\alpha$. It is to be noted that $\alpha =0$ case corresponds to $5$-dimensional Einstein analogue of EGB gravity. It turns out that the coupling constant $\alpha$ in EGB gravity has non-negligible effects on the physical quantities such as energy-density and pressure of the star. To illustrate this, we have plugged in $G$ and $c$ in appropriate places and determined the central density and pressure. In EGB gravity, the central density and pressure are obtained as $1.969 \times 10^{15}$~gm~cm$^{-3}$ and $3.318 \times 10^{35}$~dyne~cm$^{-2}$ , respectively. In its $5$-dimensional Einstein analogue the corresponding quantities take values $1.934 \times 10^{15}$~ gm~cm$^{-3}$ and $2.9 \times 10^{35}$~dyne~cm$^{-2}$, respectively. The higher curvature correction term apparently can accommodate more mass within a volume element. Most interestingly, the radius seems to increase marginally in EGB gravity as compared to its $5$-dimensional Einstein gravity.

To conclude, though many features like stability, causality and energy conditions remain unaffected, incorporation of a positive higher derivative term $\alpha$ appears to have non-negligible effect on the mass and radius of a star which provides an alternative mechanism to explain the compactness of a large class of observed pulsars.

\end{document}